\documentclass[slac_one]{revtex4}
\usepackage{graphicx}
\usepackage{fancyhdr}
\renewcommand{\baselinestretch}{1.2}
\newcommand{\pt}{\rm p_{\rm T}}
\newcommand{\ptjet}{ p_{\rm T}^{\rm jet}}

\newcommand{\ptjetcor}{p_{\rm T, cor}^{\rm jet}}

\newcommand{\yjet}{y^{\rm jet}}
\newcommand{\ptjetcal}{p_{\rm T,cal}^{\rm jet}}

\newcommand{\yjetcal}{y^{\rm jet}_{\rm cal}}

\def\njet{N_{\rm jet}}                   
\def\sigmanjet{\sigma_{N_{\rm jet}}}

\def\Zee{\mbox{$Z/\gamma^* (\rightarrow e^+ e^-)$}}
\def\Ztt{\mbox{$Z/\gamma^* (\rightarrow \tau^+ \tau^-)$}}

\def\tt{\mbox{$t \overline{t}$}}

\begin{document}


\title{
Measurement of Inclusive Jet Cross Sections in $\Zee$+jets Production  in 
{\boldmath $p\overline{p}$} Collisions at 
{\boldmath $\sqrt{s}$} = 1.96 TeV 
} 

%

\author{M. D'Onofrio, M. Mart\'\i nez, O. Salt\'o (for the CDF collaboration)}
\affiliation{IFAE/ICREA, Institut de F\'\i sica d'Altes Energies, Barcelona, E-08193, Spain.}

\begin{abstract}
Inclusive jet cross sections in $Z/\gamma^*$ events, with $Z/\gamma^*$ decaying 
into an electron-positron pair, are measured as a function of jet transverse momentum 
and jet multiplicity  in $p \overline{p}$ collisions at $\sqrt{s} = 1.96 \ {\rm TeV}$ with the upgraded Collider Detector 
at Fermilab in Run II, based on an integrated luminosity of $2.5 \ \rm fb^{-1}$.
The measurements cover the rapidity region
$| \yjet | < 2.1$
and the transverse momentum range \mbox{$\ptjet > 30$~GeV/c}.
Next-to-leading order perturbative QCD predictions are in good agreement with the 
measured cross sections.
\end{abstract}

\maketitle

\thispagestyle{fancy}

\section{INTRODUCTION}

The study of the production of electroweak bosons in association with jets of hadrons in the final
state   constitutes a fundamental item in the high-$\pt$ physics program at the Tevatron. These events
are main backgrounds to many interesting physics processes like, for example, top production, the search
for the SM Higgs, and supersymmetry. The CDF experiment has recently published  precise measurements on jets 
in events with a Z/$\gamma^*$ boson in the final state~\cite{cdfZjets}, where inclusive jet cross  sections as 
a function of jet transverse momentum and  jet multiplicity are measured and compared to pQCD predictions~\cite{mcfm}. 
At the leading order (LO) in pQCD, $ Z/\gamma^*$+jet events are driven by the processes
$gq \rightarrow  Z/\gamma^* + q$ and $q\overline{q} \rightarrow  Z/\gamma^* + g$, while higher orders
contributions, including additional parton radiation, produce mutiple jets in the final state.
Next-to-leading
order (NLO) pQCD predictions for $Z/\gamma^*$+jets production are only available for
jet multiplicities $\njet$ up to $\njet = 2$.
This contribution presents updated results with an increased data sample corresponding to a total integrated luminosity 
of $2.5 \ \rm fb^{-1}$.

\section{EVENT SELECTION}

The events are required to have two electrons with $E_T^e > 25$~GeV and a
reconstructed invariant mass in the range $66 < M_{ee} < 116$~GeV/$c^2$ around the $Z$ boson mass.  
The electron candidates are reconstructed using criteria described in~\cite{ecuts}.
In this study, one electron is required to be central ($|\eta^e|<1$) and fulfill tight selection cuts,
while the second electron is required to pass a looser selection  and to be either central 
or forward  with $1.2 < |\eta^e| < 2.8$. The events are selected to have a reconstructed primary vertex with $z$-position within
60~cm around the nominal interaction point, and at least one jet with transverse momentum
$\ptjet > 30$~GeV/$c$, rapidity in the range $|\yjetcal|<2.1$, and  $\Delta R_{e-jet} > 0.7$, 
where $\Delta R_{e-jet}$ denotes the distance between the jet and each of the two  electrons in the final state.
The main backgrounds to the $\Zee$+jets  sample arise from inclusive-jets and $W$+jets events, and
are estimated from the data. 
Other background  contributions from $\tt$, $\Zee + \gamma$, $WW$, $WZ$, $ZZ$,
and $\Ztt$+jets final states are estimated using Monte Carlo samples.
The total background in  inclusive $\Zee$+jets production
is about 12$\%$ for $\njet \geq 1$, and increases up to about 17$\%$ for $\njet \geq 3$.

\section{UNFOLDING PROCEDURE}

The measured cross sections are  corrected for acceptance and smearing effects  back to the hadron level
using {\sc pythia-tune~a}  Monte Carlo event samples~\cite{pythia,underlying}, CTEQ5L~\cite{cteq5l} parton distribution functions (PDFs) 
for the proton and antiproton, and a bin-by-bin
unfolding procedure that also accounts for the efficiency of
the $\Zee$ selection criteria.
The final results refer to hadron level
jets with $\ptjet > 30$~GeV/$c$ and  $|\yjet|<2.1$, in
a limited and well-defined kinematic range for the $Z/\gamma^*$ decay products:
$E^e_T > 25$~GeV, $|\eta^{e1}| < 1.0$, $|\eta^{e2}|<1.0$ or
$1.2 < |\eta^{e2}|<2.8$, $66 < M_{ee} < 116$~GeV/$c^2$, and  $\Delta R_{e-jet} > 0.7$.
In order to avoid any bias on the correction factors due to the particular PDF set used,
which translates into slightly different simulated $\ptjetcal$ distributions,
the {\sc pythia-tune~a}  Monte Carlo event sample is  re-weighted until it accurately follows
the measured $\ptjetcal$ spectra.
The unfolding factors
$U(\ptjetcor) = \frac{\rm d \sigma}{d \ptjet}/\frac{\rm d \sigma}{\rm d \ptjetcor}$
are computed separately for the different
measurements and vary between
2.0 at low~$\ptjet$ and 2.3 at high~$\ptjet$.

\section{COMPARISON WITH PQCD PREDICTIONS}

\begin{figure}
\includegraphics[width=70mm]{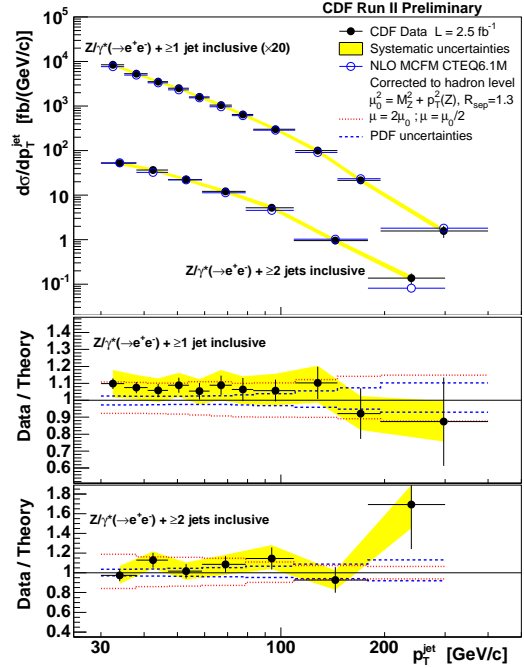}
\caption{(top)Measured inclusive jet differential cross section
as a function of $\ptjet$
(black dots) in $\Zee$+jets  with
$\njet \geq 1,2$ compared to NLO pQCD predictions (open circles).
For clarity, the measurement for $\njet \geq 1$ is scaled up by ($\times$20).
The shaded bands show the total systematic uncertainty,
except for the 5.8$\%$ luminosity uncertainty.
(middle and bottom) Data/theory ratio  as a function of $\ptjet$ for $\njet \geq 1$ and $\njet \geq 2$, respectively.
The dashed and dotted lines indicate the PDF uncertainty and the variation with $\mu$ of
the NLO pQCD predictions, respectively.
}
\label{fig:pt}
\end{figure}

Figure~\ref{fig:pt}(top) shows the measured inclusive jet differential cross sections  as a function
of $\ptjet$ in $\Zee$+jets production, with $\njet \geq 1$ and $\njet \geq 2$,
compared to NLO pQCD predictions.
The cross sections decrease by more than three orders of magnitude as $\ptjet$ increases
from  30~GeV/c up to about 300~GeV/c. The NLO pQCD predictions are computed using the
{\sc mcfm} program~\cite{mcfm} with CTEQ6.1M PDFs~\cite{cteq},  with the
renormalization and factorization scales  set to $\mu^2  =  M_Z^2 + p^2_T(Z)$, and using a midpoint~\cite{midpoint}
algorithm with $R=0.7$ and $R_{\rm sep} = 1.3$~\cite{rsep} to reconstruct jets at the parton level.
The theoretical predictions include  parton-to-hadron
correction  factors $C_{\rm had}(\njet, \ptjet)$  that approximately account for
non-perturbative contributions from  the underlying event
and fragmentation into hadrons.
In each measurement $C_{\rm had}$ is estimated using the {\sc pythia-tune~a} Monte Carlo samples, as the ratio between the
nominal $\ptjet$ distribution and the one obtained
by turning off both the interactions between proton and antiproton remnants and the
string fragmentation in the Monte Carlo samples.  The correction decreases as $\ptjet$ increases from
about 1.2 (1.26) at $\ptjet$ of 30~GeV/c  to
1.02 (1.01) for  $\ptjet > 200$~GeV/$c$ for $\njet \geq 1$ ($\njet \geq 2$), and is
dominated by the underlying event contribution.
The uncertainty on $C_{\rm had}$ is about
$10 \%$ ($17\%$) at low $\ptjet$ and goes down to $1 \%$ at high $\ptjet$
for $\njet \geq 1$ ($\njet \geq 2$).
The ratios  between data and theory as a function of $\ptjet$ are also shown in Fig.~\ref{fig:pt}.
Good agreement is observed between the
measured cross sections and the nominal theoretical predictions. 

\begin{figure}[h]
\includegraphics[width=70mm]{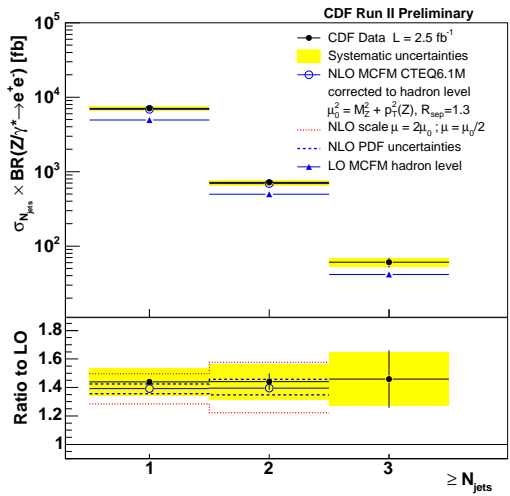}
\caption{ (top) Measured  cross section for inclusive jet production in $\Zee$ events as a function of $\njet$ compared
to LO and NLO pQCD predictions.
The shaded bands show the total systematic uncertainty,
except for the 5.8$\%$ luminosity uncertainty.
(bottom)
Ratio of data and NLO to LO pQCD predictions versus  $\njet$.
The dashed and dotted lines indicate the PDF uncertainty and the variation with $\mu$ of
the NLO pQCD predictions, respectively.
}
\label{fig:njet}
\end{figure}

Finally, Fig.~\ref{fig:njet} shows the cross sections $\sigmanjet$ for $\Zee$+jets events
up to \mbox{$\njet \geq  3$}. 
The data are compared to LO and NLO pQCD predictions. The
parton-to-hadron non-perturbative corrections vary between 1.1 and 1.4 as $\njet$ increases.
The LO pQCD predictions underestimate the  measured cross sections by a factor about 1.4 approximately independent of $\njet$.
Good agreement is observed between data and NLO pQCD
predictions. 

The final data sample collected by the CDF experiment in Run~II, where more than 6~fb${}^{-1}$ are expected,
 will make possible  further detailed studies of the event topologies. 

\begin {acknowledgments}
We would like to thank the organizers for their kind invitation to the conference, 
the wonderful atmosphere, and the discussions that they made possible.
\end{acknowledgments}

\end{document}